\documentclass[prb,twocolumn,showpacs]{revtex4-1}
\pdfoutput=1
\usepackage{graphicx}
\usepackage{times}
\usepackage{bm}
\usepackage{amsmath, verbatim}

\def\lsim{\mathrel{\rlap{\lower4pt\hbox{\hskip1pt$\sim$}}
    \raise1pt\hbox{$<$}}}                % less than or approx. symbol
\def\gsim{\mathrel{\rlap{\lower4pt\hbox{\hskip1pt$\sim$}}
    \raise1pt\hbox{$>$}}}                % greater than or approx. symbol

\def\Tr{{\text{Tr}}\,}

\def\be{\begin{equation}}
\def\ee{\end{equation}}
\def\bea{\begin{eqnarray}}
\def\eea{\end{eqnarray}}
\def\bse{\begin{subequations}}
\def\ese{\end{subequations}}

\def\Tr{{\text{Tr}}\,}

\def\be{\begin{eqnarray}}
\def\ee{\end{eqnarray}}

\newcommand{\expect}[1]{\langle #1 \rangle}

\begin{document}

\title{Amplitude mode  of the $d$-density wave state and its relevance to high-$T_c$ cuprates}
\author{Jay D. Sau$^{1}$}
\author{Ipsita Mandal$^{2}$}
\author{Sumanta Tewari$^{3}$}
\author{Sudip Chakravarty$^{2}$}

\affiliation{$^1$Department of Physics, Harvard University, Cambridge, MA 02138\\
$^2$Department of Physics and Astronomy, University of California Los Angeles, Los Angeles, California 90095-1547\\
$^3$Department of Physics and Astronomy, Clemson University, Clemson, SC
29634}

\begin{abstract}
We calculate the spectrum of the amplitude mode, the analog of the Higgs mode in high energy physics, for the $d$-density wave (DDW)  state proposed  to describe the pseudogap phase of the high $T_c$ cuprates.
Even though  the state breaks   translational symmetry by a lattice spacing and is described by a particle-hole singlet order parameter at the wave vector $q=Q=(\pi,\pi)$, remarkably, we find that the amplitude mode spectrum  can have  peaks
at both  $q=(0,0)$ and $q=Q=(\pi,\pi)$; we shall lattice spacing to unity. In general, the spectrum is non-universal, and,
depending on the microscopic parameters, can have one or two peaks in the Brillouin zone, signifying existence of two kinds of magnetic excitations.
Our theory sheds important light on how multiple
inelastic neutron peaks at different wave vectors can, in principle,  arise even with an order parameter that condenses at  $Q=(\pi,\pi)$.
\end{abstract}

\maketitle

\section{Introduction}

Ever since the discovery of  pseudogap in high temperature superconductors, it has been a profound mystery.~\cite{Norman:2005} To this day, its origin is vigorously debated. In one view, pseudogap is a remnant of the $d$-wave superconducting gap that defines a crossover temperature $T^{*}$ in the phase diagram. The other view argues for a broken symmetry at $T^{*}$. The precise nature of the broken symmetry is  debated, however.~\cite{Varma:1997,Chakravarty:2001,Kivelson:2003,Hosur:2012,Orenstein:2012,Chakravarty:2013} The central defining property of the pseudo gap is a strongly momentum dependent gap of $d_{x^{2}-y^{2}}$ character. At least at the level of  Hartree-Fock mean field theory, and perhaps even on more general grounds, it is difficult to see how  translationally invariant order parameters~\cite{Varma:1997,Oganesyan:2001} in the {\em particle-hole} channel can produce a  gap. By contrast in the {\em particle-particle} channel, i.e. in a superconductor, the gap is tied to the Fermi surface, not to the lattice, and there is no necessity for a broken translational symmetry to produce a momentum dependent gap. Here, we shall assume that much of the  phenomenology associated
with the pseudogap can be described in a unified manner by the single assumption of
a spin singlet $d_{x^2-y^2}$ density wave (DDW)~ \cite{Chakravarty:2001} and deduce some observable consequences that should help its detection.

The purpose of the present paper is to calculate the spectrum of the amplitude
mode, the analog of the recently discovered Higgs mode in high energy physics, for the DDW state. 
%related  to the inelastic neutron scattering experiments in the pseudogap phase of the high $T_c$ superconductors \cite{Li:2012,Mook:2012}. 
As we shall see, this turns out to be very unusual and was missed in a very early paper \cite{Tewari:2002} on this subject because of the simplicity of the formalism, which
did not take into account all possible fluctuations even in the context of an order parameter formalism. Nonetheless, it was shown explicitly  that a damped amplitude mode 
at $q=(\pi,\pi)$ should be observable in inelastic neutron scattering measurements. Since then it was discovered that DDW model need to be extended in order to solve the 
specific heat puzzle: to date no specific heat singularity is observed at the pseudogap transition. This extension was described in Ref.~\onlinecite{Chakravarty:2002}. 
With this extension no specific heat singularity arises at the psudogap/DDW transition. While the Hartree-Fock theory captures the broad overall picture of the phase diagram,~\cite{Nayak:2002} 
 by its very nature it cannot properly address the fluctuation spectrum. Given the recent advances in experiments, it is therefore  important to explore the consequences of the special fluctuation spectrum of the extended DDW model, which goes well beyond the Hartree-Fock theory. This is precisely what we would like to accomplish in the present paper.

 Within the extended DDW model, we shall find that although the DDW  order parameter condenses at the wave vector $q=Q=(\pi,\pi)$, remarkably,  the amplitude or Higgs mode spectrum  can be peaked
at both  $q=(0,0)$ and $q=Q=(\pi,\pi)$. These results provide important clues regarding the recent experiments on the cuprates \cite{Li:2012,Mook:2012}, which quite
unexpectedly find multiple magnetic excitation modes at different wave vectors, the precise origin of which has been mysterious.
  As to the elastic signature of singlet DDW, two neutron scattering measurements  provide some evidence for it~\cite{Mook:2002,*Mook:2004}. We believe that the identification of the DDW state can  be considerably strengthened  by careful experimentation of  the predictions of the  inelastic amplitude  spectrum that we offer.

The plan of the paper is as follows: in Sec. II we set out the model and in Sec. III we calculate the fluctuation spectrum in detail. The results are discussed in Sec. IV followed by discussion in Sec. V.

\section{The extended DDW model}
\begin{figure}
\centering
\includegraphics[width=\linewidth]{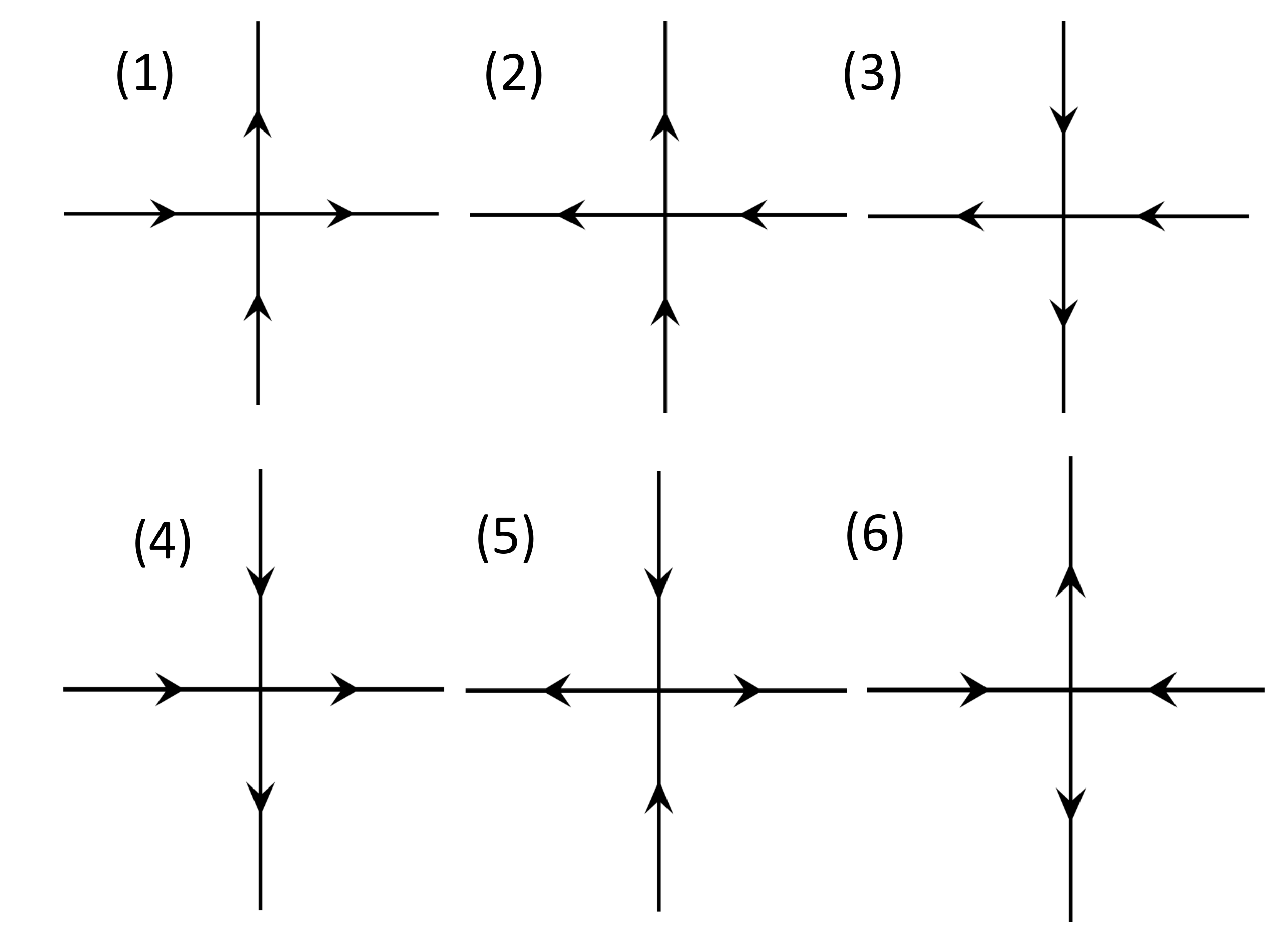}
\caption{
The six possible current vertices in the $6$-vertex model.
The vertices (5) and (6) are the AF
vertices which lead to the DDW phase with local orbital moments, while the
rest of the vertices lead to longer range current fluctuations.
}\label{Fig1}
\end{figure}

The singlet  DDW, as originally envisioned, consists  of circulating  currents,~\cite{Nayak:2000} alternating
between clockwise and anticlockwise directions in the neighboring plaquettes of
an underlying square lattice in two-dimensional ($2D$) CuO$_2$ planes. 
The particle-hole spin-singlet DDW order parameter is
\begin{equation}
\langle c_{\mathbf{k},\alpha}^{\dagger}c_{\mathbf{k+Q},\beta} \rangle = i \Delta_{Q}f_{\mathbf{k}}\delta_{\alpha\beta},
\label{eq:Order-Parameter}
\end{equation}
where $c,c^{\dagger}$ are electron annihilation and creation operators, and $\alpha,\beta$ are the spin indices; $\Delta_Q$ is the magnitude of the
order parameter and the form factor $f_{\mathbf{k}}=(\cos k_x - \cos k_y)$;  we  set the lattice constant to unity throughout.
Viewed from this perspective, the ordered state can be constructed
as a juxtaposition of two kinds of current conserving vertices, as shown in vertices 5 and 6 of Fig.~\ref{Fig1}, 
The ordered state breaks time reversal,
rotation by $\pi/2$,  parity, and translational symmetry by one lattice spacing, but the product of any two of these symmetries is preserved. The statistical mechanics belongs to the Ising universality class,~\cite{Nayak:2000} 
and in Hartree-Fock theory the order disappears when the magnitude of the bond currents vanishes
with increase of temperature. 

\begin{figure}[htb]
\centerline{\includegraphics[width=\linewidth]{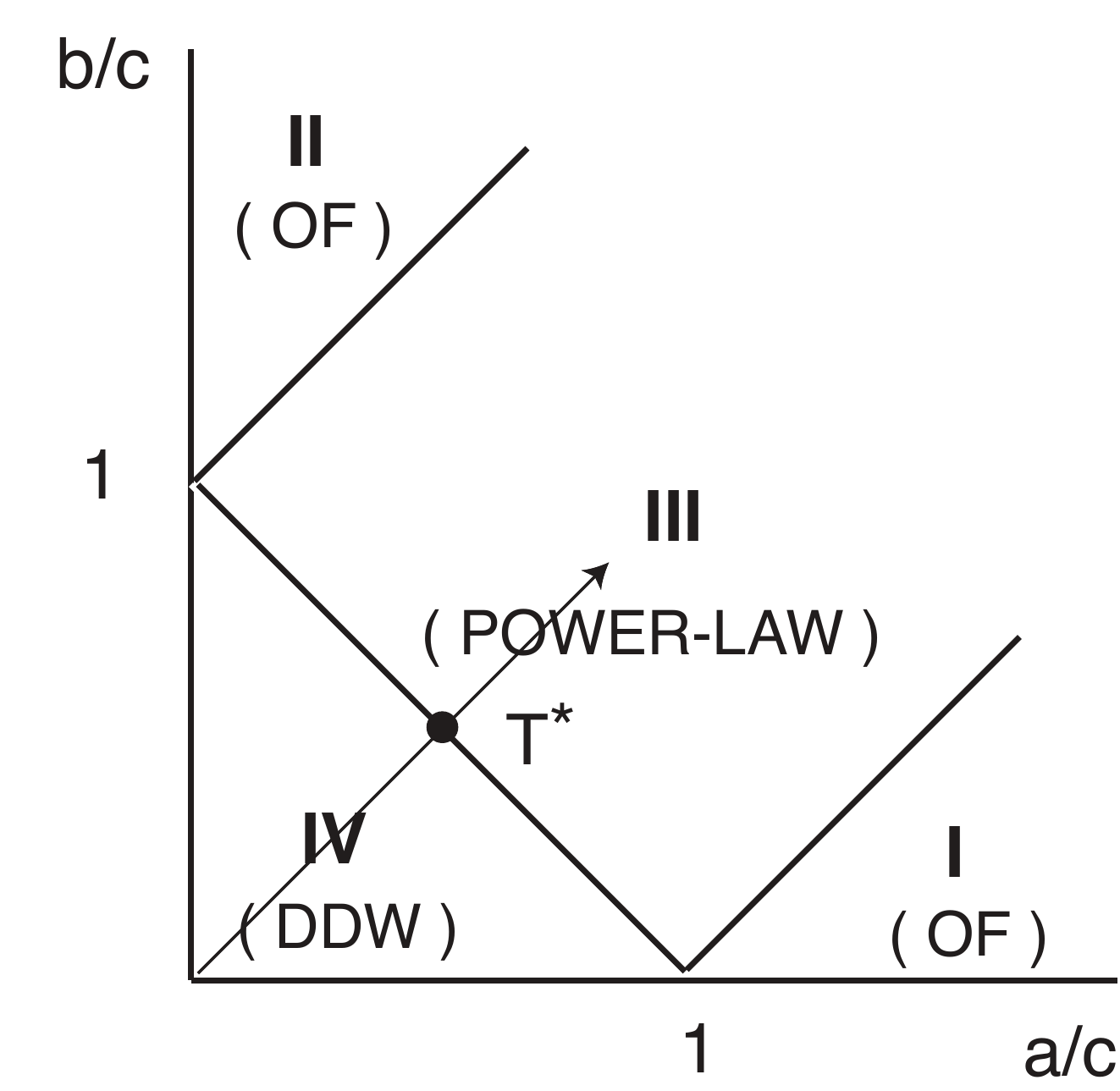}}
\caption{The phase diagram of the six-vertex model from Ref.~\onlinecite{Chakravarty:2002}; $a$, $b$ and $c$ are the three independent vertex weights discussed there The regions I and II are orbital ferromagnets (OF). The region III is the
power-law phase and the region IV is the orbital antiferromagnet with DDW order. The arrow marks a path (tetragonal
symmetry assumed) from the low temperature to the high temperature phase with the pseudogap transition at
$T^*$.}
\label{fig:phase}
\end{figure}

However,  the model was extended~\cite{Chakravarty:2002} to include fluctuations that can reverse an arrow if it is possible to do so in a current conserving manner. This can be done by enlarging the configuration space by adding four additional vertices shown  as 1, 2, 3, and 4 in Fig.~\ref{Fig1}.~\cite{Chakravarty:2002} The vertices corresponding to sources and sinks should have a large negative chemical potential. The model then belongs to the universality class of the classical six-vertex model.~\cite{Chakravarty:2002} Including the sources or sinks will convert the problem to the eight-vertex model and the concomitant spacific heat singularity at the pseudogap transition, which is not observed. The statistical mechanics of $2D$ classical six-vertex is a solved problem; the phase diagram is shown in Fig.~\ref{fig:phase}. But the dynamics is, to our knowledge, unexplored. It is indeed the collective dynamics of the model when described in the context of the electronic model appropriate to high temperature superconductors  that we wish to study here. 

 Given that the order parameter condenses at the wave vector $Q=(\pi,\pi)$,
one would have naively expected that the amplitude collective mode, the analog of the $U(1)$ Anderson-Higgs mode,   but in the particle-hole channel, would be peaked at the same wave vector.
An important result of our  paper is that the spectra of the amplitude fluctuations of the DDW state are \textit{not} confined to only the ordering wave vector $Q$, but can have finite-frequency peaks at both $q=(0,0)\equiv 0$ and $q=Q$, as well as considerable spectral weight over a substantial region of the momentum space.

In passing we note that the quantum six-vertex model described elsewhere~\cite{Syljuasen:2006} can be cast in terms of a  continuum action of a $n$-vector $(n=2$) model with cubic anisotropy.
The precise relation of the parameters in this continuum model to the parameters in the quantum six-vertex model is complex, but this mapping is not without merit. Of course at finite temperature it is an effective ``Ginzburg-Landau'' model that complements the picture in terms of the vertex model.
The isotropic $n=2$ vector model with cubic anisotropy has been extensively studied.~\cite{Rudnick:1978} Here we do not pursue this route, but pursue  the more microscopic vertex model, which is more convenient to describe the electronic excitations.

\section{Collective mode analysis of the DDW state}
\subsection{DDW Hamiltonian:}
To calculate the DDW collective mode spectrum we start with the electronic Hamiltonian which has the form,
\begin{equation}
H=\sum_{k,\sigma}(e_k-\mu)c_{k,\sigma}^\dagger c_{k,\sigma}+g^{-1}\sum_q\hat{\Delta}_q^\dagger\hat{\Delta}_q ,
\end{equation}
where
\begin{equation}
\hat{\Delta}_{q}=\frac{g}{2}\sum_k f_{k+(q-Q)/2} c_{k}^\dagger c_{k+q},\label{eq:Deltaq}
\end{equation}
Here we adopt a commonly used  dispersion $e_{k}$  based on  local density approximation~\cite{Pavarini:2001}, which is
\begin{equation}
\begin{split}
e_k&= -2t(\cos k_x + \cos k_y)+ 4t'\cos
k_x \cos k_y\\&- 2t''(\cos 2k_x + \cos 2k_y).
\end{split}
\end{equation}
For subsequent  notational simplicity, we  define the functions,
$\varepsilon_{k}=\frac{1}{2}(e_{k}+e_{k+Q})$, and
$\epsilon_{k}=\frac{1}{2}(e_{k}-e_{k+Q}).$
The band parameters are chosen to be
$t~=~0.15$ eV, $t'~=~0.3t$, and $t''~=~0.5t'$.  The  difference with the conventional LDA band structure is a rough renormalization of $t$ (from $0.38$ eV to $0.15$ eV), which is supported by experiments involving  angle resolved photoemission spectroscopy~\cite{Damascelli:2003}.
Note that in Eq.~\ref{eq:Deltaq}, as elsewhere, we shall drop the spin-index $\sigma$, as it will play no important role.

\subsection{Six vertex modes of the DDW state}
For an analysis of the collective modes  of the DDW state at a general $q$, we
have generalized the  gap parameter in Eq.~\ref{eq:Deltaq}. In this form, $\hat{\Delta}_{q}^{\dagger}=\hat{\Delta}_{-q}$, for an arbitrary $q$, which implies that the gap parameter is purely real in the real space. This constraint is necessary to ensure that no gapless phase mode is generated in the collective mode spectrum of the twofold commensurate DDW state which breaks only discrete symmetries \cite{Tewari:2002}.
From Eq.~\ref{eq:Deltaq} it follows that at $q=Q$, $\hat{\Delta}_Q$ represents the conventional DDW gap parameter shown in Fig.~\ref{Fig1} in its vertex representation, while $q=0$
leads to a gap parameter $\hat{\Delta}_{0}$ which represents uniform current flow along the $+x$ and $+y$ directions.
Choosing $q=\mathcal{Q}=Q+\bar{Q}$, where $\bar{Q}=(\pi,-\pi)$, we find currents flowing along the $+x$ and $-y$ directions. These, together with reversal of  currents, give the remaining vertices in Fig.~\ref{Fig1}. Finally, at $q=\bar{Q}$, $\hat{\Delta}_q$ allows a
gap parameter that breaks local current conservation with sites
having a current source at one vertex and a current sink in the neighboring vertex. However, in this paper we will choose the
fugacity of these current-conservation violating vertices, which are controlled by the coupling constant $g$ in front of the form factors,
to vanish~\cite{Chakravarty:2002}. Thus the operator $\hat{\Delta}_q$ represents
 the full set of current
vertices of the six-vertex model.
  We  find that only the DDW gap parameter
$\hat{\Delta}_Q$ develops a mean-field expectation value in the saddle-point solution, though we will find fluctuations from all the
other vertices as well.

Anticipating a twofold commensurate DDW order with the order parameter given by Eq.~(\ref{eq:Order-Parameter}), we first fold the full $2D$ Brillouin zone (BZ) to the reduced BZ (RBZ).
The reduced zone is defined in terms of the
rotated coordinates,
$
({k_x+k_y})/{\sqrt{2}}= k_x',
({k_x-k_y})/{\sqrt{2}}= k_y',
$
so that in  the RBZ
$k_x',k_y'\in[-{\pi}/{\sqrt{2}},{\pi}/{\sqrt{2}}]$.
Note that the Dirac points in the spectrum occur at $(k_x',k_y')=({\pi}/{\sqrt{2}},0)$ and  $(k_x',k_y')=(0,{\pi}/{\sqrt{2}})$.
The vector $Q$ is at $(\sqrt{2} \pi,0)$, while $\bar{Q}=(\pi,-\pi)$ in the original basis is now at $(0,\sqrt{2}\pi)$.

To facilitate our discussion,
we  introduce the spinor notation
$\hat{\Psi}_{k}^{\dagger}=(c^{\dagger}_{k},c^{\dagger}_{k+Q})$.
Defining a BZ periodic function $u_k$ which is 1 inside the RBZ and zero outside, $\hat{\Delta}_q$ may be written as ($\sigma_{i}, i=1,2,3$ are the conventional Pauli matrices)
\begin{equation}
\hat{\Delta}_q=\frac{g}{2}\sum_k u_k f_{k+(q-Q)/2}\Psi_k^\dagger[u_{k+q}\sigma_3\Psi_{k+q}+i u_{k+q+Q}\sigma_2\Psi_{k+q+Q}]
\end{equation}
The above expression for $\hat{\Delta}_q$ is defined for $q$ in the  full BZ. In the RBZ this corresponds to
four different gap parameters.
 It is convenient to split $\hat{\Delta}_q$ as
\begin{align}
&\hat{\Delta}_q=\frac{g}{2}\sum_k u_k u_{k+q}f_{k-Q/2+q/2}\Psi_k^\dagger\sigma_3\Psi_{k+q},\nonumber\\
&\hat{\Delta}_{q+Q}=\frac{g}{2}\sum_k u_k u_{k+q}f_{k+q/2}\Psi_k^\dagger\sigma_2\Psi_{k+q},\nonumber\\
&\hat{\Delta}_{q+\mathcal{Q}}=\frac{g}{2}\sum_k u_k u_{k+q}f_{k-\bar{Q}/2+q/2}\Psi_k^\dagger\sigma_3\Psi_{k+q},\nonumber\\
&\hat{\Delta}_{q+Q+\mathcal{Q}}=\frac{g}{2}\sum_k u_k u_{k+q}f_{k+q/2+\mathcal{Q}/2}\Psi_k^\dagger\sigma_2\Psi_{k+q}\label{hatDelta},
\end{align}
where $\mathcal{Q}=(Q+\bar{Q})$ and we have dropped the imaginary $i$ in $\hat{\Delta}_{q+Q}$ and $\hat{\Delta}_{q+Q+\mathcal{Q}}$
After this transformation, the vector $q$ only takes values which are the differences of the wave-vectors in the RBZ.
The explicit forms for these structure factors are $f^{(0)}_{0,k}=2\cos{\frac{k_x}{\sqrt{2}}}\sin{\frac{k_y}{\sqrt{2}}}$, $f^{(0)}_{Q,k}=-2\sin{\frac{k_x}{\sqrt{2}}}\sin{\frac{k_y}{\sqrt{2}}}$, $f^{(1)}_{0,k}=2\sin{\frac{k_x}{\sqrt{2}}}\cos{\frac{k_y}{\sqrt{2}}}$, and $f^{(1)}_{Q,k}=-2\cos{\frac{k_x}{\sqrt{2}}}\cos{\frac{k_y}{\sqrt{2}}}$). Considering these form factors in real space we note that
 these 4 operators (with positive and negative signs) now correspond to the 8 current orderings of the
eight-vertex model. As discussed above,  we will choose the $g$ associated with the non-current conserving vertices
to vanish, thus leaving us with a six-vertex model. The phase transition of the six-vertex model does not have a singular specific heat at its transition, whereas the eight vertex model does; this is in accord with experiments in high temperature superconductors.

\subsection{Effective action for the six vertex fluctuations}
For compactness of notation we will write $\hat{\Delta}^{(a)}_{b,q}=\hat{\Delta}_{q+b+a\mathcal{Q}}$, where $a=0,1$ and
$b=0,Q$. With this notation, the hermiticity condition on $\hat{\Delta}_q$ translates into the condition
$\hat{\Delta}^{\dagger (a)}_{b,q}=\hat{\Delta}^{(a)}_{b,-q}.$
Furthermore it is convenient to write $\hat{\Delta}^{(a)}_{b,q}=\sum_k \Psi_{k+q}^\dagger \rho^{(a)}_{b,k+q/2}\Psi_k$, where $\rho^{(a)}_{b,k+q/2}$
 are $2\times 2$ matrix-valued
structure factors.
The Hamiltonian in the spinor notation is
 \begin{equation}
H=\sum_{k}\Psi^\dagger_{k}((-\mu+\varepsilon_k)\sigma_0+\epsilon_k\sigma_3)\Psi_{k}
+ g^{-1}\sum_{q,a,b}\Big[\hat{\Delta}^{(a)\dagger}_{b,q}\hat{\Delta}^{(a)}_{b,q}\Big],
\label{eq:H-Spinor}
\end{equation}
where in the sums over $a$ and $b$, $a=1$ and $b=Q$ are dropped because these vertices do not conserve currents.
To compute the collective response of the ground state of $H$, we write the
partition function for $H$ as a path integral over an imaginary time action~\cite{Popov:1987}.
In this formalism, the four-fermion term $\sum_{q,a,b}\hat{\Delta}_{b,q}^{(a)\dagger}\hat{\Delta}_{b,q}^{(a)}$ in Eq.~\ref{eq:H-Spinor} can be decoupled using Hubbard-Stratonovich-transformation by introducing auxiliary real fields $\Delta^{(a)}_{b,q,\omega}$ (i.e. satisfying the constraint
$\Delta_{b,-q,-\omega}^{(a)}=\Delta^{*(a)}_{b,q,\omega}\,,$ corresponding to their operator counterparts) as
\begin{align}
& \exp{ \Big[ g^{-1}\sum_{q,\omega,a,b} \hat{\Delta}^{(a)\dagger}_{b,q}\hat{\Delta}^{(a)}_{b,q} \Big]} \nonumber\\
&= \int D\Delta^{(a)}_{b,q,\omega} \exp{ \Big[  g^{-1}\sum_{q,\omega,a,b} \Big(2 \hat{\Delta}_{b,q,\omega}^{(a)\dagger}\Delta_{b,q,\omega}^{(a)} - \Delta_{b,q,\omega}^{*(a)}\Delta_{b,q,\omega}^{(a)} \Big) \Big]}.
\end{align}
This leads to the effective action,
 \begin{align}
&S=\sum_{k,\omega}\Psi^\dagger_{k,\omega}((-i\omega-\mu+\varepsilon_k)\sigma_0+\epsilon_k\sigma_3)\Psi_{k,\omega}\nonumber\\
&- 2 \, g^{-1}\sum_{q,\omega,a,b}\hat{\Delta}_{b,q,\omega}^{(a)\dagger}\Delta_{b,q,\omega}^{(a)}+g^{-1}\sum_{q,\omega,a,b}\Delta_{b,q,\omega}^{*(a)}\Delta_{b,q,\omega}^{(a)}.
\label{eq:Effective-Action}
\end{align}
%corresponding to their operator counterparts in Eq.~\ref{eq:constraint}.
Expressing the $\hat{\Delta}$ operators in Eq.~(\ref{eq:Effective-Action}) in terms of the fermion spinors
 $\Psi$'s and then performing the Grassmannian path integral over the quadratic  terms
leads to the effective action,
\begin{align}
&S=g^{-1}\sum_{q,\omega,a,b}\Delta_{b,q,\omega}^{*(a)}\Delta_{b,q,\omega}^{(a)}\nonumber\\
&-\Tr[\ln(M_{0,k_1,\omega_1}\delta_{k_1,k_2}\delta_{\omega_1,\omega_2}+\delta M^{k_1,k_2}_{\omega_1,\omega_2})]\label{eq:S2},
\end{align}
where
 \begin{align}
&M_{0,k,\omega}=(-i\omega-\mu+\varepsilon_k)\sigma_0+\epsilon_k\sigma_3- \sum_{a,b}\rho^{(a)}_{b,k}\Delta^{(a)}_{b,0}\label{M0},\\
&\delta M^{k_1,k_2}_{\omega_1,\omega_2}=-u_{k_1}u_{k_2}\sum_{a,b}\rho^{(a)}_{b,(k_1+k_2)/2}\Delta^{(a)}_{b,k_1-k_2,\omega_1-\omega_2}.
\end{align}

\subsection{Saddle point treatment of the six vertex fluctuations}
The mean-field equation for the DDW state is obtained by extremizing the action $S$  by differentiating Eq.~\ref{eq:S2} with respect to $\Delta_{k,\omega}$ and setting the derivatives to zero.
This leads to the equation,
\begin{align}
&\frac{1}{g}= \sum_{k} \frac{ f_k^2}{ E_k} [n_F(\varepsilon_k+E_k-\mu)-n_F(\varepsilon_k-E_k-\mu)].\label{eq:DeltaQ}
\end{align}
for the order parameter  $\Delta_Q$;  $E_k=\sqrt{\epsilon_k^2+  f_k^2\Delta_Q^2}$
 and $n_F(E)$ is the Fermi-function.
The chemical potential $\mu$ is determined from the hole-doping $x$.
There are no static saddle point solutions corresponding to $\expect{\hat{\Delta}_{q}}$ and $\expect{\hat{\Delta}_{q+\mathcal{Q}}}$ for the range of parameters
considered.  However, interestingly, the collective mode spectrum for the DDW
state will contain a component also near $q\sim 0$, which can be interpreted as orbital ferromagnetic fluctuations of the bond currents.

To compute the response to a perturbation $\zeta$ that couples linearly
to $\Delta$ through a term in the action $S_{ext}=\int dqd\omega \zeta_{q,\omega}\Delta_{q,-\omega}$ (where we have suppressed the indices $a,b$)
one needs to extremize the action in the presence of such a perturbation so that $\delta (S+S_{ext})/\delta \Delta=0$. This leads to an equation
$\delta\Delta=-(\delta^2 S/\delta^2\Delta)^{-1}\zeta$. Thus the response kernel $V$ is given by the inverse of $\delta^2 S/\delta\Delta^2$.
Restoring all the indices, the second derivative of the action $S$ in   Eq.~\ref{eq:S2} is obtained by considering the second order terms
 in  $\Delta_{q,\omega}$ in Eq.~\ref{eq:S2} as
\begin{align}
&\delta S^{(2)}=   \frac{1}{2} \sum_{k_1,\omega_1} \nonumber\\
&\Tr[u_{k_1}u_{k_1+q}\{\sum_{a,b}\rho^{(a)}_{b,k_1+q/2}\Delta^{(a)*}_{b,q,\omega_1-\omega_2}\}M_{0,k_1+q,\omega_1+\omega}^{-1}\nonumber\\
&\{\sum_{a,b}\rho^{(a)}_{b,k_1+q/2}\Delta^{(a)}_{b,q,\omega_1-\omega_2}\}M_{0,k_1,\omega_1}^{-1}]\nonumber\\
& + g^{-1}\sum_{\omega,a,b}\Delta_{b,q,\omega}^{*(a)}\Delta_{b,q,\omega}^{(a)}.\label{deltaS}
\end{align}
Performing the Matsubara summations  and analytically continuing to real frequency leads to
\begin{align}
&\delta S^{(2)}=\sum_{a,b,a',b'}\Delta_{b,q,\omega}^{(a)*}U^{(a,a')}_{b b'}(q,\omega)\Delta^{(a')}_{b',q,\omega},
\end{align}
where $U^{(a,a')}_{b b'}(q,\omega)=\delta^2 S/\delta\Delta^{(a)*}_{b,q,\omega}\delta \Delta^{(a')}_{b',q,\omega}$ are given by
\begin{align}
&U^{(a,a')}_{b b'}(q,\omega) = \sum_{k_1,m,n}\Lambda_{m,n}(k_1,k_1+q;\omega),  \,\nonumber\\
& \Tr( \rho^{(a)}_{b,k_1+q/2} A_{m,k_1} \rho^{(a')}_{b',k_1+q/2} A_{n,k_1+q}) + g^{-1}\delta_{a,a'}\delta_{b,b'},\label{eq:U}
\end{align}
and
\begin{align}
&\Lambda_{m,n}(k_1,k_1+q;\omega)=\nonumber\\
&-\frac{n_F(m E_{k_1}+\varepsilon_{k_1}-\mu)-n_F(n E_{k_{1}+q}+\varepsilon_{k_1+q}-\mu)}{2\,(\omega-m E_{k_1}+n E_{k_1+q}-\varepsilon_{k_1}+\varepsilon_{k_1+q}+i\delta)},
\end{align}
with
$A_{m,k}=\frac{1}{2 E_k} ( E_k + m \epsilon_k\sigma_3 -  m f_k\Delta_Q\sigma_2 )$.
For the rest of the paper we will focus on the spectra at the special $q$ points $q=Q$ and $q=0$, where the $U$ coefficients take the form
\begin{align}
&U^{(a,a)}_{Q,Q}(0,\omega)=\sum_{k:E_{k}>|\mu-\varepsilon_{k}|}\frac{\omega^2 f_{k}^2-4 \Delta_Q^2 f_{k}^4-4\epsilon_k^2 (f_{k}^2-f^{(a)2}_{Q,k})}{ E_{k}(\omega^2-4 E_{k}^2)},\label{eq:resQ}
\end{align}
and
\begin{align}
&U^{(a,a)}_{0,0}(0,\omega)=\sum_{k:E_{k}>|\mu-\varepsilon_{k}|}f_k^2\frac{\omega^2 -4 \epsilon_k^2-4\Delta_Q^2 (f_{k}^2-f^{(a)2}_{0,k}) }{ E_{k}(\omega^2-4 E_{k}^2)},\label{eq:res0}
\end{align}
respectively
at $T=0$.
In this limit, we find that all terms, where $b\neq b'$ or $a\neq a'$,  vanish
because of the different symmetries of the form factors under $k_x\rightarrow -k_x$ and $k_y\rightarrow -k_y$.
% The expressions for general $T$ and $q$ are more complicated and will not be explicitly given here.

The measurement of dissipation from any probe that couples to the fields $\hat{\Delta}_q$ and $\hat{\Delta}_{q+Q}$ is related to the
imaginary part of the response, which would be of the form
\begin{align}
&\Im[\expect{\Delta^{(a)}_{b,q,\omega}\Delta^{(a')}_{b',-q,-\omega}}]=\Im[V^{(a,a')}_{b,b'}(q,\omega)],\label{eq:corr0}
\end{align}
with
\begin{align}
&V(q,\omega)=\left(U^{(a,a')}_{b,b'}(q,\omega)\right)^{-1}\label{M},
\end{align}
where $(a,b)$ and $(a',b')$ are treated as matrix indices.
\begin{comment}
Peaks in the response function thus correspond to poles of $V$.
We conjecture (NEEDS TO BE CHECKED) that the on-site current-current correlator near $q\sim (0,0)$ is given by $V_{11}$ and the correlator near
$q\sim(\pi,\pi)$ is given by $V_{22}$.
\end{comment}
Thus the peaks in the imaginary part of the spectrum near $q=0$ and $Q$ are given by the zeroes of the functions in Eq.~\ref{eq:resQ} and Eq.~\ref{eq:res0}.
The $a=0$ (corresponding to the conventional DDW order parameter) propagator near $Q$,  given by Eq.~\ref{eq:resQ},
 can  be easily seen to vanish when $\omega\sim 2 \Delta_{Q}$ because of the numerator.
 On the other hand, low $\omega$ zeroes in the other propagators only emerge when $\mu$ is such that the denominator $(\omega^2-4 E_k^2)$ can vanish while
satisfying the condition $E_k>|\mu-\varepsilon_k|$. This occurs in the presence of a
 non-zero next nearest neighbor coupling $t'$, which breaks particle-hole symmetry in the band structure.
\section{Results}
 There is no mean field OF order
coexisting with the twofold  commensurate DDW order in the regime of doping studied here.
\begin{figure}
\centering
\includegraphics[width=\linewidth]{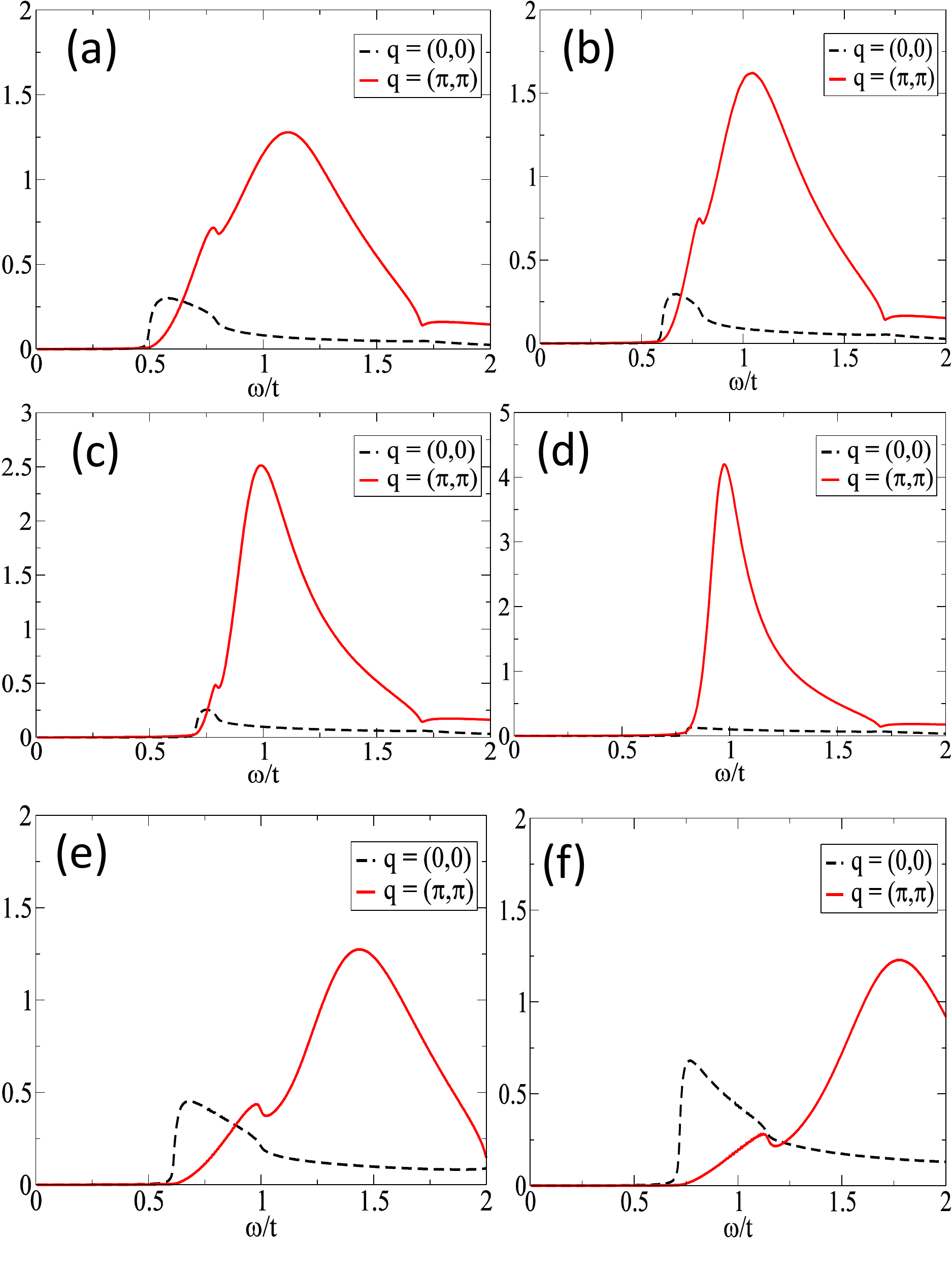}
\caption{
(Color online) Collective mode response of the commensurate DDW state as a function of $\omega/t$ for $q=Q$ (solid curve) and $q=0$ (dashed curve)
for the band parameters given  in the text. In the  panels (a)-(c),  $\Delta_Q=0.3 t$, and  $x=0.06$ (a),  $x=0.1$ (b), $x=0.14$ (c), $x=0.18$ (d).
The emergence of an inelastic $q=0$ fluctuation peak with underdoping is a signature of hidden underlying OF fluctuations co-existing with the commensurate DDW  fluctuations centered at $q=Q$. The vertical axes correspond to arbitrary units, but are the same for all panels.The panels (e) and (f) correspond to $x=0.06$ and $\Delta_Q=0.4 t$ and $\Delta_Q=0.5 t$ respectively and demonstrates further the non-universal aspects  as a function of parameters.}
\label{Fig2}
\end{figure}
In contrast, from  Fig.~\ref{Fig2}  it is clear that in the underdoped regime there is a noticeable
finite-frequency peak at $q=0$ coexisting with
a  finite frequency peak at $Q$. The intensity at $q=0$  goes down
with increasing $x$, but the  amplitudes  strongly depend on  the microscopic parameters, which implies that
  the amplitude fluctuation spectra can have finite frequency peaks at both wave vectors $q=0$ and $q=Q$, or only a single peak. Such non-universality, commonly unrecognized, indicates that different families of cuprates, or even samples at different hole doping within the same family of cuprates, may have different fluctuation spectra. 
 The non-universality of the collective modes and the coexistence of finite frequency peaks at multiple wave vectors are the central
 aspects of our  work. These results are direct consequences of the vertex model framework of the fluctuations for the DDW state and may have important consequences for the inelastic neutron scattering experiments. 
 We also note previous collective mode calculations in the framework of $t-J$-model, but with important differences with our present results.~\cite{Greco:2006}

 In principle,
one can also expect fluctuations associated with the $q=0$ nematic state, which, as shown in Ref.~[\onlinecite{Kee:2008}], is comparable in energy to the DDW state. This nematic order parameter $\Delta_{\text{nem}}= \sum_{k} f_{k}c^{\dagger}_{k}c_{k}$ is a close cousin of the order parameter discussed here. Finally, because the orbital magnetic moment contribution to inelastic neutron scattering is expected to be enhanced at long wavelengths,  the relatively small peak at $q=0$  can produce scattering
peaks comparable to that at $Q$. However, the detailed kinematics addressing quantitatively the neutron scattering signatures of the amplitude modes  will be meaningful once the experiments provide a more detailed picture.
\section{Conclusions}
In summary, we calculate the spectrum of the amplitude mode, the analog of the $U(1)$ Higgs mode, for the commensurate DDW state proposed earlier to explain the anomalous phenomenology
of the pseudogap phase of the high-$T_c$ cuprates. 
To properly take account of the fluctuations of the DDW state we map the DDW fluctuation problem on to the fluctuations of a
vertex model, which is the most natural description of the DDW current fluctuations.
Our central result is that the amplitude or Higgs mode spectrum of the DDW state can be peaked at both wave vectors
 $q=(0,0)$ and $q=Q=(\pi, \pi)$, even though the ordered state condenses only at the wave vector $Q$. 
 
 The emergence of a $q=0$ peak, even with a mean field state that breaks the lattice translation symmetry (such as the twofold commensurate DDW) indicates
 $q=0$ fluctuations are hidden within  the DDW state. We find that these  fluctuations reflect an {\em orbital ferromagnetic} (OF)  phase in the six-vertex generalization of the model, as mentioned above.  In its phase diagram one finds OF as well as orbital antiferromagnetic (DDW) phases; see Fig.~\ref{fig:phase}. The existence of such striking $q=0$ fluctuations, even in the absence of an OF order parameter,
can be important in interpreting  inelastic neutron scattering experiments.~\cite{Li:2012} However,  because the response is at  higher frequencies, they are subject to considerable degree of non-universality and serves as a warning that if our theory is correct, experimental signatures should not be unique across materials, as well as within a given material, as a function of doping, frequency, etc.

 We provide predictions as well as  shed important light on the recent
neutron scattering experiments \cite{Li:2012,Mook:2012}, which evince multiple magnetic excitations at different wave vectors in the pseudogap phase, the origin of which has remained  mysterious within the existing theoretical framework of the the pseudogap regime of the cuprates.

\begin{acknowledgments}
J. D. S acknowledges support from the Harvard Quantum Optics Center.  I. M. was funded by  David. S. Saxon Presidential Term Chair at UCLA. S. T. would like to thank DARPA-MTO, Grant No. FA9550-10-1-0497 and NSF, Grant No. PHY-1104527 for support. S. C. was supported by a grant from  NSF, Grant No. DMR-1004520.
  J. D. S. and S. C. also acknowledges support from NSF Grant No. PHY-1066293 and the  hospitality of the Aspen Center for Physics where the work was completed.
 We would also like to thank R. B. Laughlin, B. Keimer, M. Greven and H. A. Mook for discussions. We particularly thank H. A. Mook for sharing with us unpublished work of him and his coworkers.
\end{acknowledgments}

%%%%%%%%%%%%%%%%%%%%%%%%%%%%%%%%%%%%%%%%%%%%%%%%%%%%%%%%%%%%%%%%%%%%%
%%%%%%%%%%%%%%%%%%%%%%%%%%%%%%%%%%%%%%%%%%%%%%%%%%%%%%%%%%%%%%%%%%%%%

%%%%%%%%%%%%%%%%%%%%%%%%%%%%%%%%%%%%%%%%%%%%%%%%%%%%%%%%%%%%%%%

\end{document}